\documentclass[a4, twocolumn]{article}
\usepackage[utf8]{inputenc}
\usepackage{mathtools}
\usepackage{amsmath,amssymb}
\usepackage{braket}
\usepackage{amsmath}
\usepackage{bm}
\usepackage{authblk}
\usepackage{babel}
\usepackage{graphicx}
\usepackage[right=1.5 cm, left=1.5 cm, top=2cm]{geometry}

\title{Superradiant Thomson scattering from graphite in the extreme ultraviolet}

\date{}
\author[a,b]{Claudia Fasolato}
\author[a]{Elena Stellino} 
\author[c]{Emiliano Principi}
\author[c]{Riccardo Mincigrucci}
\author[c]{Jacopo Stefano Pelli-Cresi}
\author[c]{Laura Foglia}
\author[d]{Paolo Postorino}
\author[a,e]{Francesco Sacchetti}
\author[a,f]{Caterina Petrillo}

\affil[a]{Universit\`a di Perugia, Dipartimento di Fisica e Geologia, I-06123 Perugia, Italy}
\affil[b]{CNR-ISC, Istituto dei Sistemi Complessi, I-00185 Roma, Italy}
\affil[c]{Elettra-Sincrotrone Trieste SCpA, I-34149 Basovizza, Trieste, Italy}
\affil[d]{Sapienza Universit\`a di Roma, Dipartimento di Fisica, I-00185 Roma, Italy}
\affil[e]{CNR-IOM, Istituto Officina dei Materiali, I-06123 Perugia, Italy}
\affil[f]{AREA Science Park, I-34149 Padriciano, Trieste, Italy}

\begin{document}
\twocolumn[
  \begin{@twocolumnfalse}
    \maketitle
    %\begin{abstract}
      \noindent We study the Thomson scattering from highly oriented pyrolitic graphite excited by the extreme ultraviolet, coherent pulses of FERMI free electron laser (FEL). An apparent nonlinear behavior is observed and fully described in terms of the coherent nature of both exciting FEL beam and scattered radiation, producing an intensity dependent enhancement of the Thomson scattering cross section. The process closely resembles the Dicke's superradiant phenomenon  and also triggers the generation of coherent, low-\textit{q} ($<$ 0.3 \AA$^{-1}$), low energy phonons. The experimental data and analysis provide quantitative information on the sample characteristics, absorption, scattering factor and coherent phonon energies and populations, and open the route for the investigation of the deep nature of complex materials.
      \\\\
    %\end{abstract}
  \end{@twocolumnfalse}
       ]      

\section*{Introduction}
Radiation-matter interaction can be generally described using low order perturbation theory, due to the relatively small coupling constant $\alpha = e^2 / \hbar c \simeq 1/137$ \cite{heitler,xray-theory}. That approximation is considered accurate, in the non-relativistic range, when the incident and scattered radiation can be represented as singly occupied photon states.
That condition breaks down in exotic environments, \textit{e.g.} high energy density astrophysical objects, like pulsars \cite{astro1, astro2}, where the radiation states take the form of coherent states with large ($>1$) average occupation numbers. A (Compton) scattering theory involving coherent radiation states has been long available \cite{coherence-X}, but specific applications to laboratory scattering experiments do not exist yet.

In the last decades, Free Electron Lasers (FELs) opened new routes to study different states of matter \cite{SLAC1, FLASH, SLAC-Xray}, extending Thomson scattering at short wavelengths (0.05-0.2 nm) to the study of low $Z$ elements out of equilibrium \cite{PRL-Xray, thom3, thom4, thom5, quo-vadis}, in the pump-probe scheme \cite{thom-rev, ppscattering, eis}. In this way, the electron gas at high temperatures ($k_B T \approx E_F$) was probed in different materials, while the temporal evolution of pump-triggered, coherent modes at low energy ($\approx$ meV)  could be observed \cite{coherent-phon1,coherent-phon2, LiF}. 

Here, we exploit the high degree of coherence, ensured by the seeded nature of FERMI FEL (Trieste, Italy), and its high power density ($\approx$ TW/cm$^2$) in a small space-time range ($\lesssim 10^{-6}$ cm$^2 \times 10^2$ fs) \cite{LiF, FERMI}, to explore the nonlinearities in radiation-matter interaction by a properly designed scattering experiment, where coherence in quantum mechanics (QM) plays a crucial role. Over the last decade, the linear response approximation was observed to break down already in transmission experiments, employing intense and short ($< 100$ fs) extreme ultraviolet (EUV) pulses \cite{abs-sat1, abs-sat2, abs-sat3}. There, the nonlinearity observed was ascribed to the abrupt FEL-induced modification of the sample state, not fundamentally affecting the probe-sample coupling. Instead, in our scattering experiment, we observe a nonlinear amplification of the cross section which is \textit{intrinsic}, \textit{i.e.} not primarily associated with a FEL-modified sample state.

In scattering processes, the presence of a coherent state of incoming radiation has no specific effect on the probe-sample coupling \cite{coherence-X}, while a coherent final state produces an {\it amplification} of the scattering cross section at the lowest order of the perturbation theory. 
That effect strongly relates to Dicke's theoretical prediction for spontaneous radiation processes \cite{dicke} and is in line with the description of superradiant processes by Gross and Haroche \cite{superradiant2} in the form: {\it "the superradiant emission is a cooperative process involving in a collective mode all the atoms of the sample"}. 
That kind of phenomenon is hardly visible in the scattering by electrons, where the final states are limited by conservation rules only, while they are more likely observed in phonon scattering, owing to the upper bound of phonon energy, limiting the number of available scattering channels. Hence, a sub-ps duration of the FEL probe pulse (much shorter than the typical phonon lifetimes \cite{phonon-lifetime}, yet much longer than the radiation period $\lambda_0/c \approx 0.01$ fs) can trigger a coherent scattering process with a final coherent photon state in the Glauber's sense \cite{glauber}, resulting in the observation of a fundamentally nonlinear process.

To date, the observation of superradiant scattering has been reported mainly in astrophysical contexts, or analogues \cite{Torres2017}, and in the study of the collective excitation of cold atom condensates \cite{Inouye1999}. Superradiant effects in condensed matter have been observed in emission processes from cooperatively interacting excited electron states \cite{Cong2016,Masson2022}. We demonstrate here that a new range of superradiant scattering experiments in condensed matter is possible, exploring the scattering from phonons with final coherent radiation states.

That regime is achievable by employing EUV photons \cite{FERMI} so that only a small fraction of the Brillouin zone (BZ) (\textit{i.e.}, few modes) is accessible and the nonlinear process is activated. Conveniently, in stacked 2D systems, along the unique axis, a reduced number of states in the $q-$space is sampled, and the number of scattering channels is smaller: here, the hexagonal graphite is studied as a prototype for 2D systems. That class includes many other interesting materials, among which \textit{e.g.} transition metal dichalcogenides \cite{TMD}. We will discuss how the described nonlinear scattering process can trigger nonthermal populations of coherent phonons at low  momentum $q$ in such systems by employing sub-ps coherent EUV pulses. 

\section*{Experimental design}
\begin{figure}
\centering
\includegraphics[width=0.5\textwidth]{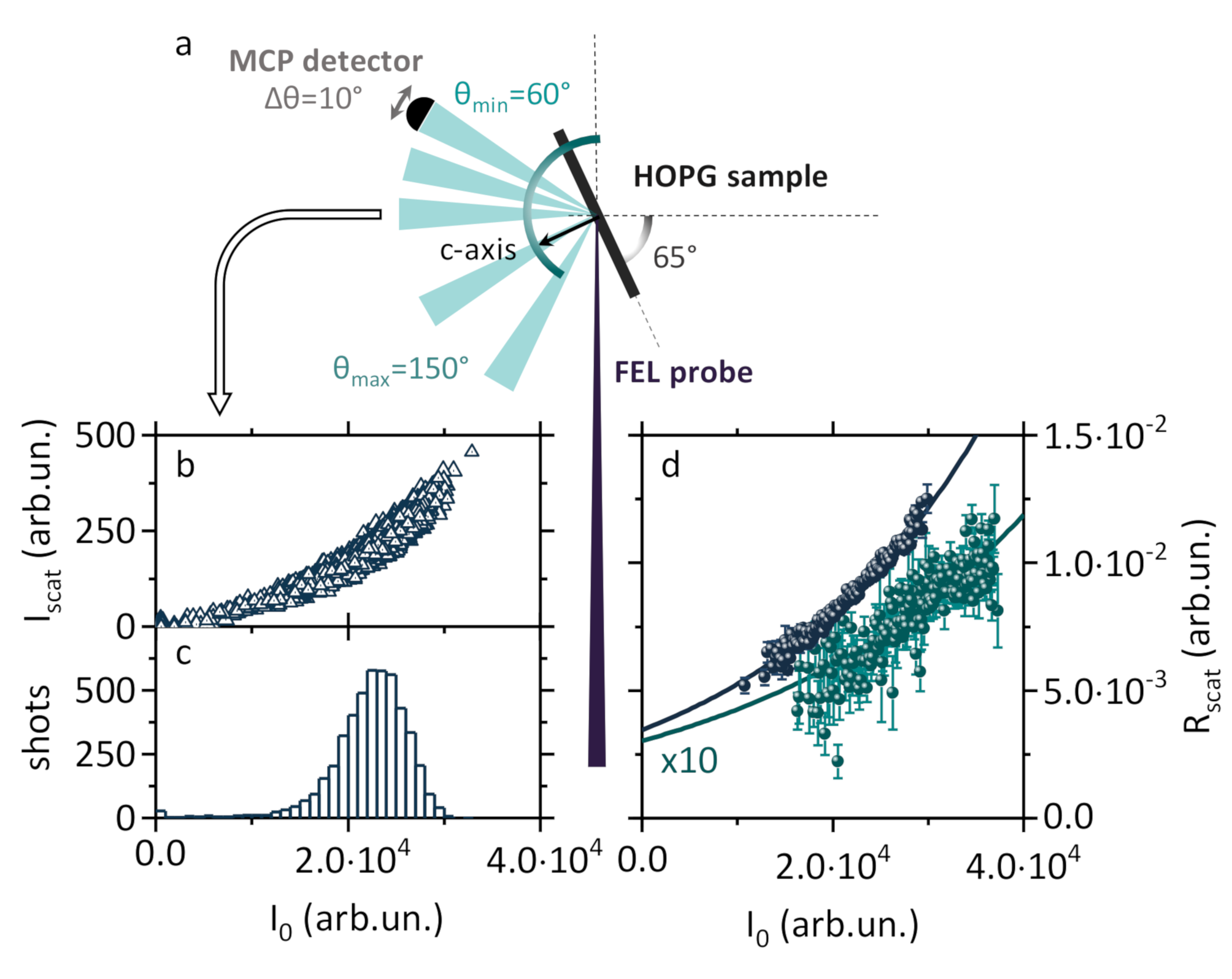}
\caption{a) Experimental setup (top view). The FEL beam impinges on the sample tilted by $65^{\circ}$. The scattered radiation is collected by the MCP, with a $10^{\circ}$ angular opening, at variable $\theta$ (60$^{\circ}-150^{\circ}$). b) Measured scattered intensity $I_{scat}$ versus FEL incident intensity $I_0$ per single shot, for $\lambda_0^{(1)} = 4.08$ nm, $\epsilon_0 = $ LV, $\theta = 90^{\circ}$. c) $I_0$ histogram of data in panel b). d) $R_{scat}$ vs $I_0$ for $\lambda_{0}^{(1)}$ (blue dots) and $\lambda_{0}^{(2)}$ (cyan dots, data multiplied by a factor 10 for better visualization), $\epsilon_0$ = LV, $\theta=90^{\circ}$. Notice the similar exponential relative increase and the remarkably different value at $I_0 = 0$. Data are corrected according to the statistical analysis (see text). Solid lines are single exponential fits to the data.}
\label{fig1_setup}
\end{figure}

We carried out a Thomson scattering study on highly oriented pyrolytic graphite (HOPG) using the $\approx$70 fs EUV pulses of FERMI \cite{FERMI2}. Two  harmonics of FEL-2 configuration were employed, $\lambda_0^{(1)} = 4.08$ nm and $\lambda_0^{(2)}= 4.74$ nm, which correspond to photon energies $\simeq \pm 20$ eV around the carbon absorption K-edge ($E_{K}\approx 285$ eV) \cite{ref353}. Such wavelengths allow to access the previously unexplored region of low $q$ transfer \cite{SLAC-Xray, PRL-Xray}. The sample orientation was fixed (to keep the scattering volume constant), the HOPG c-axis forming a 65$^{\circ}$ angle with the incident beam, while the angle of detection was varied in the $\theta =  60^{\circ} - 150^{\circ}$ range. The scattering geometry is sketched in Fig.\ref{fig1_setup}a. Two FEL linear polarizations ($\epsilon_0$, LH and LV, parallel or perpendicular to the scattering plane) were used for both $\lambda_0$. The incident intensity $I_0$ was determined, pulse by pulse, by the photocurrent from the FEL focusing mirror and a microchannel plate (MCP) detector measured the scattered intensity $I_{scat}$. The energy density was about 0.2 J/cm$^2$ with 1 $\mu$J pulse. 
In the worst case ($\lambda_0^{(1)}$), given an absorption length of about 100 nm and the 300 eV photon energy, there  is an energy release of about 0.3 eV/atom per pulse: only the accumulation of hundreds of FEL pulses could produce damage, as was directly observed. Accordingly, on each sample spot a total of 5 single-shot acquisitions were collected to guarantee the minimum sample damage. In these conditions, only the occupation number of a few, specific vibrational modes is expected to change due to the scattering event.

\section*{Results and discussion}

\subsection*{Thomson scattering data: pretreatment and observations}
Owing to the intrinsic fluctuations of the FEL lasing process (see a histogram of the incident intensity $I_0$ in Fig.\ref{fig1_setup}c), each independent dataset at given $\lambda_0,\epsilon_0, \theta$ of our experiment (Fig.\ref{fig1_setup}b) was treated by averaging the ratio of $I_{scat}$ to the incoming $I_0$, $R_{scat} (I_0) = I_{scat}(I_0) / I_0$, over several different sample points. The average was taken in small enough intervals of $I_0$. Statistically meaningful data were obtained rejecting $I_0$ intervals with too few ($<$5) data within them, and (rare) fluctuations larger than twice the standard deviations. 

In linear Thomson scattering, $R_{scat} (I_0)$ does not depend on $I_0$ resulting as an intrinsic system property, proportional to the scattering cross section $\Sigma(\lambda_0, \epsilon_0, \theta) = \int d^2 \sigma (\lambda_0, \epsilon_0, \theta) /d \Omega d E$, integrating over final energy and detector solid angle and summing the final polarizations. The observed $R_{scat}(I_0)$ versus $I_0$, shown in Fig.\ref{fig1_setup}d, appears far from a constant. That happens for every $\lambda_0, \epsilon_0$. We ruled out the possible experimental artifacts: a saturation of $I_0$ is not expected and the low scattered intensity does not support a nonlinear trend of the MCP detector. Furthermore, both $I_0$ and $I_{scat}$ show Gaussian-like fluctuation distributions, thus supporting the absence of undesired anomalies.

\subsection*{Physical model of the scattering cross section}

To derive useful information from the experiment we need a model of the Thomson cross section in the present conditions. As it is well known, Thomson scattering originates from the coupling between electrons and the radiation field, with both electronic states and nuclear vibrations contributing. Thus, the final energy-integrated cross section is the sum of the {\it electronic} $d \sigma / d \Omega \vert_{e}$ and the {\it phononic} $d \sigma / d \Omega \vert_{ph}$ contributions. 
At the first perturbative order, the cross section is factorized into two terms: a structure factor, containing $\alpha^2$ and the effect of the system structure, and the sum of the correlation functions describing the electron and phonon dynamics \cite{heitler,xray-theory}.

In the present context, initial and final radiation fields are properly described using coherent states \cite{glauber} $\ket{I}$ and $\ket{F}$, where all the occupation numbers contribute with Poisson statistics \cite{coherence-X}, instead of the photon states in the Fock space, with zero or one occupation numbers. Accordingly, one can calculate the transition matrix element $\mathcal{M}_{I, F} \, \propto \, \sum _{i l} {\bra{F} e^{-i (\bm{k} - \bm{k}_0) \cdot (\bm{r}_i + \bm{u}_l)} \ket{I}}$, where $\bm{r}_i$ and $\bm{u}_l$ are the electron position and nucleus displacement operators, and $\bm{k}_0$, $\bm{k}$ are the incident and final photon wavevectors. The coherent states $\ket{I}$ and $\ket{F}$ are eigenvectors of the photon annihilation operators with complex eigenvalues $\alpha_I$ and  $\alpha_F$ and average occupation numbers $n_I = \vert \alpha_I \vert^2$ and $n_F = \vert \alpha_F \vert^2$. The scattering cross section becomes proportional to $n_F$ and a final coherent state is expected when $n_I$ is large and the number of final channels is small.
A coherent $\ket{F}$ is not expected for $d \sigma / d \Omega \vert_{e}$, which has many open final state channels, whereas for $d \sigma / d \Omega \vert_{ph}$, the situation is different. In the present experiment $\lambda_0 \approx 4$-5 nm, hence the accessible volume in the reciprocal space is a small fraction of the first BZ, and only a few phonon modes, \textit{i.e.} final state channels, are involved. Furthermore, a coherent initial state is present. 
We have:

\begin{equation}
\nonumber
  \frac{d\sigma}{d\Omega} \bigg|_{ph} = \frac{d\sigma^{+}}{d\Omega}\bigg|_{ph}+\, \frac{d\sigma^{-}}{d\Omega}\bigg|_{ph} = \frac{k}{k_0} \left[ \, S^+_{ph} (\bm{q}) \, n^+_F + S^-_{ph} (\bm{q}) n^-_F \right]
 \end{equation}

\noindent
where we distinguish the different channels for phonon creation ($^+$) and annihilation ($^-$). 
$S^+_{ph} (\bm{q})$ and $S^-_{ph} (\bm{q})$ are the phonon static structure factors, \textit{i.e.} the integrals of the dynamic structure factors corresponding to $n_F^+$ and $n_F^-$, respectively:

\begin{equation}
\nonumber
\begin{split}
& S^+_{ph} (\bm{q}, \omega) = N_{{\bm q} j} C({\bm q} j) (\tilde n_{{\bm q} j} + 1) \delta(\omega - \omega_{{\bm q} j}) \, \delta({\bm ko} - {\bm k} + {\bm q}) \\ 
& S^-_{ph} (\bm{q}, \omega) = N_{{\bm q} j} C({\bm q} j) (\tilde n_{{\bm q} j}) \delta(\omega + \omega_{{\bm q} j}) \, \delta({\bm ko} - {\bm k} - {\bm q})
\end{split}
\end{equation}

\noindent
Here, $C(\bm{q}j) = \left| {\bm F}_{{\bm q} j} \cdot {\bm q} \right|^2 P({\bm \epsilon},{\bm \epsilon}_0) / (2 M \omega_{{\bm q} j})$ with $M$ atomic mass, and $P({\bm \epsilon},{\bm \epsilon}_0)$ accounts for the effect of the beam polarizations.  
$\bm{F}_{\bm{q}j}$, $\hbar \omega_{\bm{q}j}$ and $\tilde n_{\bm{q}j}$ are the structure factor, energy and occupation number of the phonon state of momentum $\bm{q}$ in the $j$-th branch. $N_{{\bm q} j}$ is the number of normal modes in the scattering volume defined by the scattering solid angle. Considering that $ \omega_{\bm{q}j} \ll c k_0$, we can safely set in the following $k/k_0 = 1$. 

Finally, the scattered photon rate, $dn/dt = dn^{+}/dt + dn^-/dt$, is evaluated considering a time dependent incident intensity, proportional to the measured $I_0$ through an unknown efficiency $K_0$, $n_0 f(t) = K_0 \, I_0 f(t)$, with $\int_{-\infty}^{+\infty} \, f(t) dt = 1$:
\begin{equation} \label{diff_eq}
    \begin{cases}
    \displaystyle{\frac{dn^+(t)}{dt} = n_0 \, f(t) \, N_{{\bm q} j} \, C({\bm q} j) [\tilde n_{{\bm q} j} + 1] n^+(t)
    }\\[6pt]
    \displaystyle{\frac{ d n^{-}(t)}{dt} = n_0 \, f(t) \, N_{{\bm q} j} \, C({\bm q} j) \tilde n_{\bm{q}j}(t) n^{-}(t)}\\[6pt]
    \end{cases}
\end{equation}
where $\tilde n_{\bm{q}j}(t) = \tilde n_B(\hbar\omega_{\bm{q}j})+n^+(t)-n^-(t)$ is the time dependent occupation number of the $\bm{q} j$ phonon mode and $\tilde n_B(\hbar\omega_{\bm{q}j})$ is its Bose occupation number. In Eq.[\ref{diff_eq}], it is not specified that the quantum numbers associated with creation and annihilation have wavenumbers that are slightly different (negligibly, since $\omega_{\bm{q}j} \ll c k_0$). Both channels contribute to $I_{scat}$, as do all the modes within the detector solid angle. 

Notice two important points: first, since QM governs the processes, the rate of phonon creation is higher than that of annihilation, $n^{+}(t) - n^{-}(t) > 0$ and, consequently, $\tilde n_{{\bm q} j}(t) > \tilde n_B(\hbar \omega_{{\bm q} j})$ and, second, the presence of $N_{{\bm q} j}$ times $C({\bm q} j)$ will have a determining role in the exponential growth of the ratio of scattered to incident intensities.%, increasing the latter.

\subsection*{Experimental data fitting}
In order to use the model to describe the experimental data, we recall that (i) $I_{scat}$ is measured within the collection time of the MCP detector ($t_c \approx$ 1-2 ns) and (ii) the creation and annihilation processes cannot be distinguished (the detection is energy integrated). Accordingly, we take 
$$R_{scat} = \frac{ I_{scat}}{I_0} = \frac{K_0}{K} \, lim_{t \to \infty} \left[ \frac{n(t)}{n_0} \right]$$
with the (unknown) proportionality constant related to the detection efficiencies $K_0$ and $K$. To model $\lim_{t \to \infty} \, n(t)/n_0$ we use the analytical solution for the ratio $n^+(t)/n^-(t)$, with integration constant from the system thermodynamic equilibrium state and $g(t) = \int_{-\infty}^t \, dt' f(t')$. Fixing a time $t_f$ larger than the pulse duration, we get:

\begin{equation}\label{R}
\rho (t_f) = \frac{n^{+} (t_f)}{n^{-} (t_f)} =  \frac {\tilde n_{{\bm q} j}(t_f) + 1}{\tilde n_{{\bm q} j}(t_f)} \, \exp \, [n_0 N_{{\bm q} j} C({\bm q} j) g(t_f)]
\end{equation}

\noindent
From Eqs.[\ref{diff_eq}] and [\ref{R}] we obtain the following relation, describing the exponential growth of the scattered to incoming intensity ratio.  

\begin{equation}\label{eqn:fit}
\begin{split}
\frac{n(t_f)}{n_0}  \, 
\simeq & \, C(\bm{q}j) [2\tilde n_B(\hbar\omega_{\bm{q}j})+1]  \, \cdot \\
                     & \cdot \exp\Big[ \frac{1}{2} n_0 N_{{\bm q} j} C(\bm{q}j) \Big(\tilde n_{\bm{q}j}(t_f) + \frac{\rho(t_f)}{\rho(t_f)+1} \Big) \Big]
\end{split}
\end{equation}
The factor $C(\bm{q}j)$ is larger for low energy longitudinal modes, and it is zero for transverse modes because the structure factor is proportional to the projection of the eigenvector on ${\bm q}$. In X-ray scattering in the 0.1 nm wavelength region, this contribution is fairly simple, because it is directly obtained from the phonon eigenvectors tightly related to the nuclear crystal symmetry. In the case of the scattering at relatively low energy, and at an energy close to an absorption edge, ${\bm F}_{{\bm q} j} \cdot {\bm q}$ contains much more information because the structure factor vector ${\bm F}_{{\bm q} j}$ depends on both the ground and excited electron states. That makes the experiments more demanding, but also provides new means for the investigation of complex materials and better validation of the theoretical estimates of the real and imaginary parts of the scattering factors. 
In the present case one can assume that, among the twelve modes of HOPG at each $\bm{q}$, the only ones contributing significantly are the longitudinal acoustic mode (LA, $\omega_{\bm{q}j} \simeq c_l(\hat {\bm q}) q$) and the lowest energy longitudinal optic mode (LO, $\omega_{\bm{q}j} \approx$ const.) \cite{graphite-phon-n, graphite-phon-X, graphite-phon-X2}. The LA dispersion is known to be strongly anisotropic as the sound velocity $c_l$ is $c_l^{(c)}$ = 2.38 meV nm = 3620 m/s along the unique axis and $c_l^{(a)}$ = 17.1 meV nm = 25900 m/s on the hexagonal plane, with negligible direction dependence \cite{elastic-const-graphite}. Notice that the dependence of $c_l(\hat q)$ on $\hat q$ is unknown when its direction is intermediate between the hexagonal plane and the hexagonal axis. Also notice that in $\tilde n_{\bm{q}j}(t)$ we neglect the effect of the finite phonon lifetime ($\gg 1$ ps from the measured lineshape of the LA and LO phonon modes \cite{graphite-phon-n, graphite-phon-X2}) which is much longer than the incident pulse $f(t)$ duration ($\approx 70 $ fs).

As a preliminary test of the model, in Fig.\ref{fig1_setup}d we show a single exponential fitting to the $R_{scat}$ data. The fit is adequate in the whole $I_0$ range explored (almost a factor of two) for both wavelengths. 
Besides the expected exponential trend, the most striking feature in Fig.\ref{fig1_setup}d is the change of $R_{scat}$ on varying $\lambda_0$. A scale factor of 10 is needed to make $R_{scat}$ at $\lambda_0^{(2)}$ of the same order as that at $\lambda_0^{(1)}$. 
From the instrumental point of view, we observe that the detector efficiencies lead to an overestimate of $R_{scat}(\lambda_{0}^{(1)})$ by a factor around 2.5 (about $\times 1.25$ from the MCP $I_{scat}$ detector,  $\times 1/0.5$ for the $I_0$ detector). 
Intrinsically, $R_{scat}$ depends on three ($\lambda_0$-dependent) contributions: the carbon scattering factor $\vert f_s(\lambda_{0})\vert^2$, the absorption proportional to $1/\mu(\lambda_{0})$, and the sampled region in the BZ $V_{scat}(\lambda_{0})$, so that $R_{scat} \propto |f_s|^2 V_{scat} / \mu$. 
Thus, to interpret our experimental observation, we seek an estimate of the $\lambda_0$-dependence of $|f_s|^2 $ and $\mu$. 

A first evaluation is obtained from the free atom data provided by NIST \cite{NIST}, with large uncertainties and limited applicability because of the obvious differences between HOPG and free carbon atoms. Following \cite{NIST}, $\vert f_s(\lambda_{0}^{(1)})\vert^2 \simeq 3.2 \, \vert f_s(\lambda_{0}^{(2)}) \vert^2$ (with large uncertainties) and $\mu(\lambda_{0}^{(1)}) \simeq 16.7 \, \mu(\lambda_{0}^{(2)})$, while for the sampled region in the BZ we have $V_{scat}(\lambda_{0}^{(1)}) \simeq 1.56 \, V_{scat}(\lambda_{0}^{(2)})$. 
Taking into account the different experimental sensitivities at the two $\lambda_0$, this leads to expecting  $R_{scat}(\lambda_0^{(1)}) \approx 0.80 \,  R_{scat}(\lambda_0^{(2)})$, quite far from the present observations. 

\begin{figure}
    \includegraphics[width=0.5\textwidth]{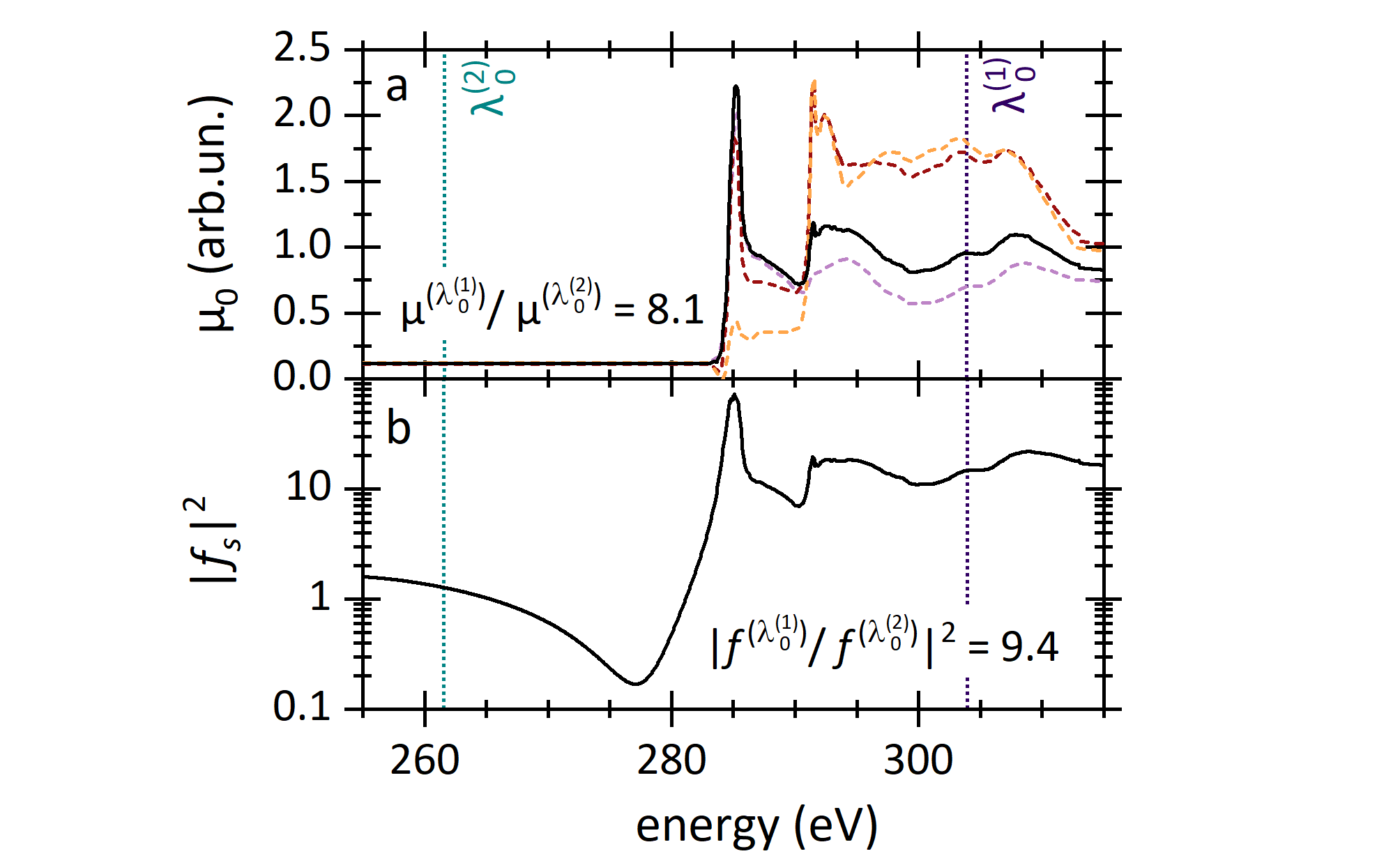}
    \caption{a) Absorption coefficient $\mu_0$ versus photon energy in the range around the K$_1$ carbon absorption edge (solid black lines). The absorption coefficient at variable scattering angle from \cite{ref353}, used for the extrapolation of $\mu_0$ in the present case, are also shown as coloured dashed lines. The vertical dotted lines indicate the photon energies corresponding to $\lambda_0^{(2)}$ and $\lambda_0^{(1)}$. b) Energy dependence of the scattering factor $|f_s|^2$, estimated from data in (a) as described in the text.}
    \label{new-fig2}
\end{figure}

A possibly more accurate estimate is based on the energy dependence of $\vert f_s(\lambda_0) \vert^2$, as determined from available experimental absorption data \cite{ref353} according to the formal relationships described in \cite{29a}. From the optical theorem, coherently with the material-specific energy dependence of $\mu(\lambda_0)$, it is possible to obtain a quantitative estimate of $\vert f_s^{\lambda_0} \vert^2$ using the simple equation
$$
f(\omega) = f_0 + f'(\omega) + i f"(\omega) = f_0 + f'(\omega) + i \frac{\omega \, \mu(\omega)}{4 \pi c r_0 n_{at}}
$$
$$
f'(\omega) = \frac{2}{\pi} \, {\cal P} \int_{-\infty}^{+\infty} \, \frac{\omega' \, f"(\omega')}{\omega'^2 - \omega^2} \, d\omega'
$$
$r_0$ being the electronic radius and $n_{at}$ the atomic density. 
The result of this calculation, carried out using the absorption data of Ref.\cite{ref353} properly interpolated to estimate $\mu (\lambda_0)$ at the present incident angle, is shown in Fig. \ref{new-fig2}. It is seen that $\vert f(\omega) \vert^2$ increases of a factor about 10 in going from $\lambda_0^{(2)} = 4.74$ nm to $\lambda_0^{(1)} = 4.08$ nm, a ratio much higher than that derived from the free atom calculations \cite{NIST}. The experimental trend of $\mu(\lambda_0)$ is also in better agreement with the higher intensity at the short wavelength, yielding $R_{scat}(\lambda_0^{(1)}) \approx 4.5 \,  R_{scat}(\lambda_0^{(2)})$, a value much closer to the experimental observations. Notice that this estimate depends on the strategy used for the calculations and the quality and spectral extension of the available $\mu (\lambda_0)$ data. 

It must be stressed that both the described \textit{a priori} evaluations of $R_{scat}(\lambda_0)$ are characterized by a limited accuracy, due to the difficulties in the experimental determination of the energy dependence of the scattering factor, as well as that of the absorption coefficient when bulk materials are considered. Here, we will demonstrate that the present study can provide solid quantitative information on the above mentioned physical quantities, otherwise hardly accessible experimentally. 

 For a quantitative analysis of $R_{scat}$, we calculate $S_{ph}(\bm{q})$ for the two phonon modes LA and LO. We add to it the low-$q$ prescription $S_e(q) \propto (\hbar / 2 m \omega_p) q^2$, $m$ being the electron mass and $\omega_p$ the plasma frequency \cite{pines}, fixed by comparing to the free atom contribution and rescaling to the phonon scattering. 
Since Eq.[\ref{eqn:fit}] contains a time dependent phonon occupation number we can use $\tilde n_{{\bm q} j}(t) - \tilde n_B(\hbar \omega_{{\bm q} j}) = n^+(t) - n^-(t)$  to describe its mild growth, because the above relations provide the closed form:

\begin{equation}
    \tilde n_{{\bm q} j}(t) = \frac{\tilde n_B(\hbar \omega_{{\bm q} j}) \, F_g(t)}{1 - n_0 N_{{\bm q} j} \, \int_{-\infty}^t \, f(\tau) F_g(\tau) d \tau}
    \label{eqn:example}
\end{equation}
with 
$
F_g(t) = \exp[n_0 N_{{\bm q} j} C({\bm q} j) \int_{-\infty}^t \, f(\tau) r(\tau) d \tau]
$, where we defined $r(t) = \rho(t)/[\rho(t) - 1]$.

The so obtained model needs two scale factors for $I_{scat}/I_0$ from LA and LO modes and one for $I_0$ because the intensities are measured in arbitrary units. The same coefficients are used for the growth of $\tilde n_{{\bm q} j}(t_f)$ with $n_0$ by means of Eq.[\ref{eqn:example}]. A fourth constant was added to describe the (unknown) $c_l(\hat {\bm q})$ change with ${\bm q}$ direction. According to the model we described, the fit was performed using the following contribution for both phonon modes involved in the scattering 
\begin{equation} 
\nonumber
\begin{split}
S_{ph}({\bm q}) =  A \, \frac{q^2}{(2 M \omega_{{\bm q} j})} & [\tilde n_B(\hbar \omega_{{\bm q} j}) + 1] \cdot \\ & \exp{ I_0 \, B_e \, A \left[\tilde n_{{\bm q} j} + \frac{\rho(t_f)}{[\rho(t_f) + 1] } \right] }
\end{split}
\end{equation}
For each phonon mode, the parameter $A$ is equal to the ratio of the two efficiency constants $K = n(t_f )/I_{scat}$ and $K_0 = n_0 /I_0$ times $C({{\bm q} j}) [2 \tilde n_B(\hbar \omega_{{\bm q} j}) + 1]$, while the common growing constant $B_e = N_{{\bm q} j} K_0$ is the same for all phonon modes.
We adopted the phonon dispersion measured in \cite{graphite-phon-n, graphite-phon-X2} since a much larger set of data would be needed to take the phonon energies as free parameters. This model well fitted each dataset, typically about 600 experimental points, at four $\lambda_0, \epsilon_0$ configurations using {\it only} 3 free parameters, dependent on the wavelength and polarization, plus just one {\it fixed} parameter to describe the $c_l$ anisotropy. That was modelled by a smooth function connecting the experimentally determined $c_l$ values along the high symmetry directions as asymptotic values. The anisotropy parameter, depending on the sample characteristics only, useful to describe this large body of data, strongly supports the present superradiant model of the coherent radiation scattering: indeed, a possible change in the sample state, produced by the incoming radiation, could not be described using a single parameter. No $q$ dependence was assumed for the phonon structure factor, a choice suggested by the small $q$ range, which is confined well within about 1/3 of the BZ width.

\begin{figure}
\centering
\includegraphics[width=\linewidth]{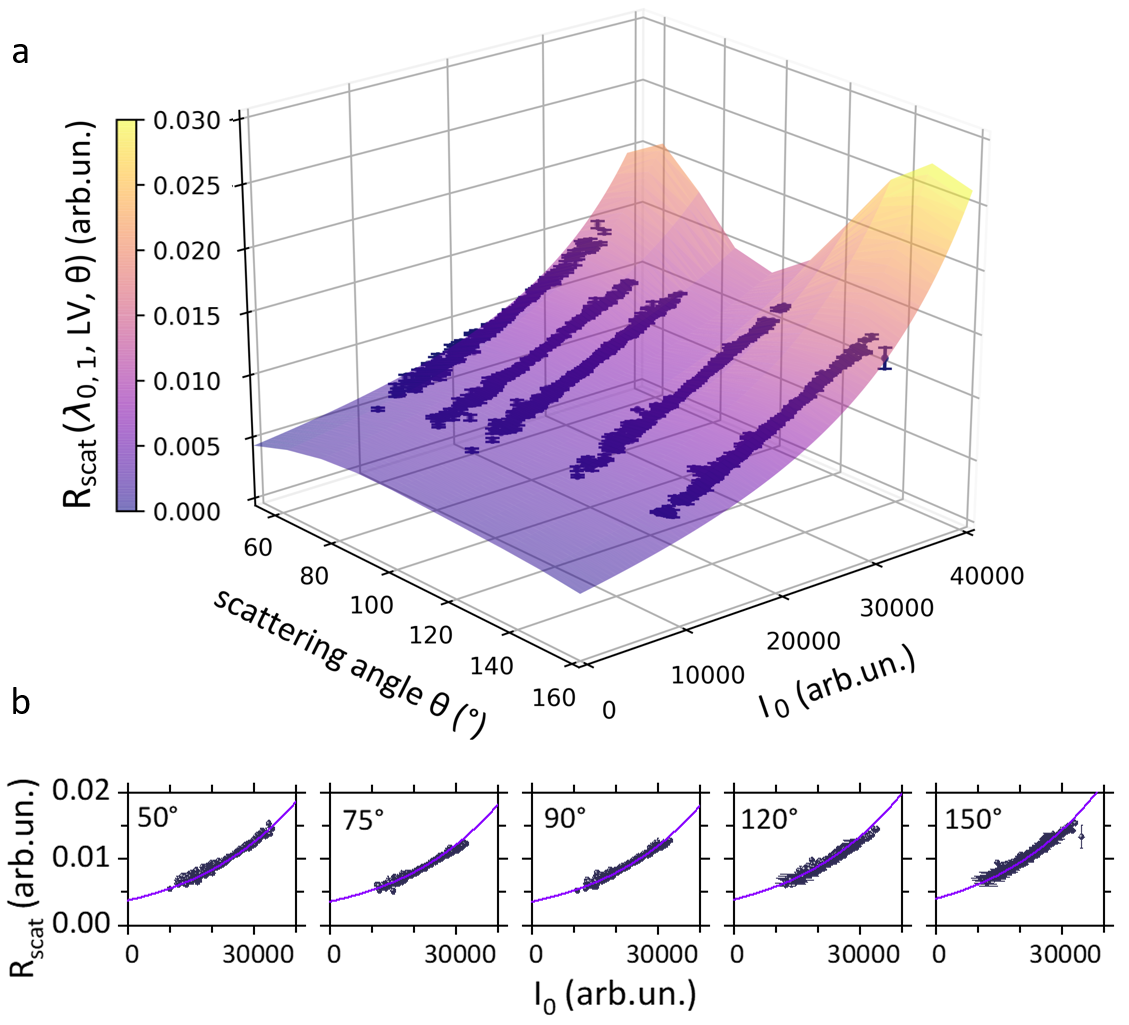}
    \caption{a) $R_{scat}$ vs $I_0$ at different $\theta$ (thus, ${\bm q}$) for $\lambda_0^{(1)} = 4.08$ nm, $\epsilon_0 =$ LV. Blue dots: experimental data; surface: data fitting.
    b) Plots of data and fitting function shown in panel a) for the specific, sampled scattering angles. Similar results are obtained for all the other $(\lambda_0, \epsilon_0)$ datasets.
    }
    \label{fig4_vs-q}
\end{figure}

\subsection*{Scattering amplitudes, phonon dispersion and coherent phonon population}

The experimental $R_{scat}$ as a function of $\theta$ and $I_0$ is shown in Fig.\ref{fig4_vs-q}, where the good agreement with the model is evident. Several quantitative information can be extracted by data fitting and will be described in the following, along with considerations, based on available data on HOPG, that strongly support our analysis and interpretation. 

The first very important result is the ratio of the growth coefficients $B_e \propto N_{{\bm q} j} \propto 1/\mu$ at 4.74 nm to that at 4.08 nm, $b_e = B_e (\lambda_0^{(2)})/B_e (\lambda_0^{(1)})$. It is found $b_{e,\mathrm{LV}} = 13.64 \pm 0.60$ and $b_{e,\mathrm{LH}} = 15.17 \pm 0.76$ with vertical and horizontal incoming polarization, respectively. These experimental results are close to the available estimate of the ratio of the linear absorption coefficients equal to 16.7 from the free atom theory \cite{NIST} and 8.1 from the experimental trends in Fig.\ref{new-fig2}. That result corroborates the model and introduces an approach for a quantitative determination of $\mu$ in bulk materials in this EUV or soft x-ray range. 

The second quantitative result is the ratio of the amplitude coefficients $A$ at the two wavelengths, $a = A(\lambda_0^{(1)})/A(\lambda_0^{(2)})$, containing $C(\bm{q}j)$ and thus information on the phonon structure factor \cite{TDS}, for the acoustic ($a_{\mathrm{l}}$) and optical ($a_{\mathrm{o}}$) modes. For vertical and horizontal polarization, the ratios are $a_{\mathrm{l, LV}} = 11.10 \pm 0.26$ and $a_{\mathrm{o,LV}} = 11.82 \pm 0.27$ and $a_{\mathrm{l, LH}} = 11.77 \pm 0.29$ and $a_{\mathrm{o, LH}} = 8.60 \pm 0.24$. These numbers are a quantitative estimate of the significant wavelength dependence of the scattering factor and polarization factor. 
From the $A$ coefficients at varying the polarization, we find that, at $\lambda_0^{(1)} =4.08$ nm, $A_{\mathrm{l,LV}}/A_{\mathrm{l,LH}} = 0.7917 \pm 0.0076$, $A_{ \mathrm{o,LV}}/A_{\mathrm{o,LH}} = 0.7606 \pm 0.0090$ and, at $\lambda_0^{(2)} =4.74$ nm, $A_{\mathrm{l,LV}}/A_{\mathrm{l,LH}} = 0.840 \pm 0.027$, $A_{ \mathrm{o,LV}}/A_{ \mathrm{o,LH}} = 0.553 \pm 0.019$. This last result indicates a rather small polarization effect on the amplitudes of the two modes. We recall that at photon energy close to the absorption edge also the polarization dependence of the scattering factors shows an energy dependence \cite{29a}.
It is relevant to notice that the present study can open the door to the determination of both real and imaginary parts of $ f_s(\lambda_0)$ in complex materials like graphite. Considering the difficulty in determining the linear absorption coefficient by transmission experiments, and the scattering factors by both experiments and theory, the observed effect opens new ways to the study of not only coherent radiation processes, but also fundamental, and hardly accessible, material properties. Notably, when coherence effects are observed, the information on $f_s$ is contained within both the multiplying coefficient and the exponent of the trend in Eq.[\ref{eqn:fit}], thus providing a more solid experimental estimate of the scattering factor, also considering that the growth coefficient $B_e$ is common to all the phonon modes involved. 

It is remarkable that, while the scattered intensity is quite lower below the absorption edge, the exponential increase as a function of the incoming intensity, and hence the effect of the coherence, remain similar. 
Indeed, the exponent factors, proportional to the product of $A$ and $B_e$ are negligibly reduced ($ < 20 \%$) at $\lambda_0^{(2)}$ in all the conditions explored. 
That is in agreement with the description of Eq.[\ref{eqn:fit}] and with what is readily observed from the trends in Fig. \ref{fig1_setup}d.

\begin{figure}
    \centering
    \includegraphics[width=\linewidth]{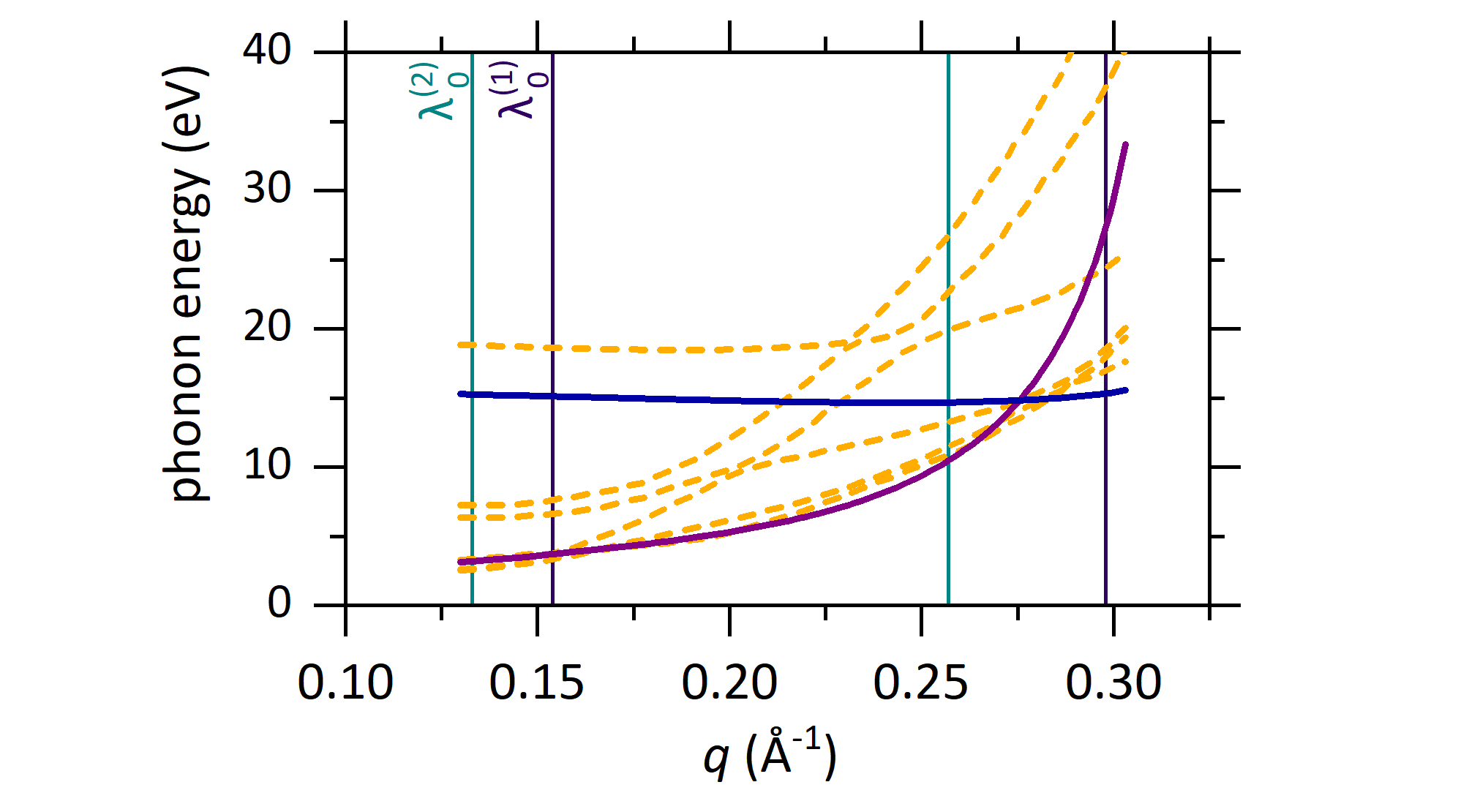}
    \caption{Phonon dispersion curves (orange dashed lines) calculated by DFT in the transferred $q$ region of interest. The vertical lines mark the $q$ range explored in the experiment with $\lambda_0^{(1)}$ (blue) and $\lambda_0^{(2)}$ (cyan) excitation. The phonon dispersion is calculated performing an average over the possible crystal orientation in HOPG (obtained by rotations about the $c$-axis). The solid lines show the dispersion curves extracted by the fitting of experimental data for the low energy acoustic (purple) and optical (blue) phonon modes.}
    \label{new-fig}
\end{figure}

From the extrapolation of $R_{scat}$ at $I_0 \rightarrow 0$, we infer that the amplitude of $S_{ph}(\bm{q})$ has a small angular dependence. That occurs because the strong anisotropy of $c_l(\hat {\bm q})$ (see the LA phonon dispersion in Fig.\ref{new-fig}) produces a decrease of the scattering contribution from the LA mode ($\approx 1/12$) at growing $\theta$, which is compensated by the $q^2$ increase of the LO part ($\approx \times 5$), an effect independent of polarization and wavelength. Some tests to check for the presence of a structure factor variation suggest that this contribution is smaller than the sensitivity of the experiment, $\approx 10$\%.

The observed variation of $c_l(\hat {\bm q})$ is not in agreement with the trend derived from the graphite elastic constants, therefore the system is not equivalent to an elastic solid, even if the crystal symmetry is taken into account \cite{elastic-const-graphite}.
To have an additional evaluation of the strong anisotropy of $c_l(\hat {\bm q})$ a density functional theory (DFT) calculation was performed using the exchange-correlation potential for the gradient approximation of ref.\cite{graphite-theory} and the all electron LAPW DFT code Elk \cite{elk-code}. The calculation has been carried out by performing an average of the dispersion relations obtained by rotation around the $c$-axis. That is necessary because of the random orientations of the small crystallites on the hexagonal plane of HOPG. 
In principle, only the longitudinal component of the phonon modes can contribute to the scattering in the first BZ. Considering the structure of graphite and the fact that, when the transferred momentum is not along a high symmetry direction the phonon polarization is neither purely longitudinal nor transverse, all the modes tend to contribute here. In Fig.\ref{new-fig}, the complex calculated dispersion curves are shown. A qualitative agreement is observed with the model of the fit which uses two major modes only. That result further corroborates both the presented description of the scattering data and its potential applicability in performing new kinds of investigations using more extended sets of experimental configurations.
The results show that the experiment can detect the anisotropic bonding of graphene layers in HOPG, while a more extended sampling of the scattering volume is needed to thoroughly assess the aforementioned smaller contributions. 

\begin{figure}
    \centering
    \includegraphics[width=.8\linewidth]{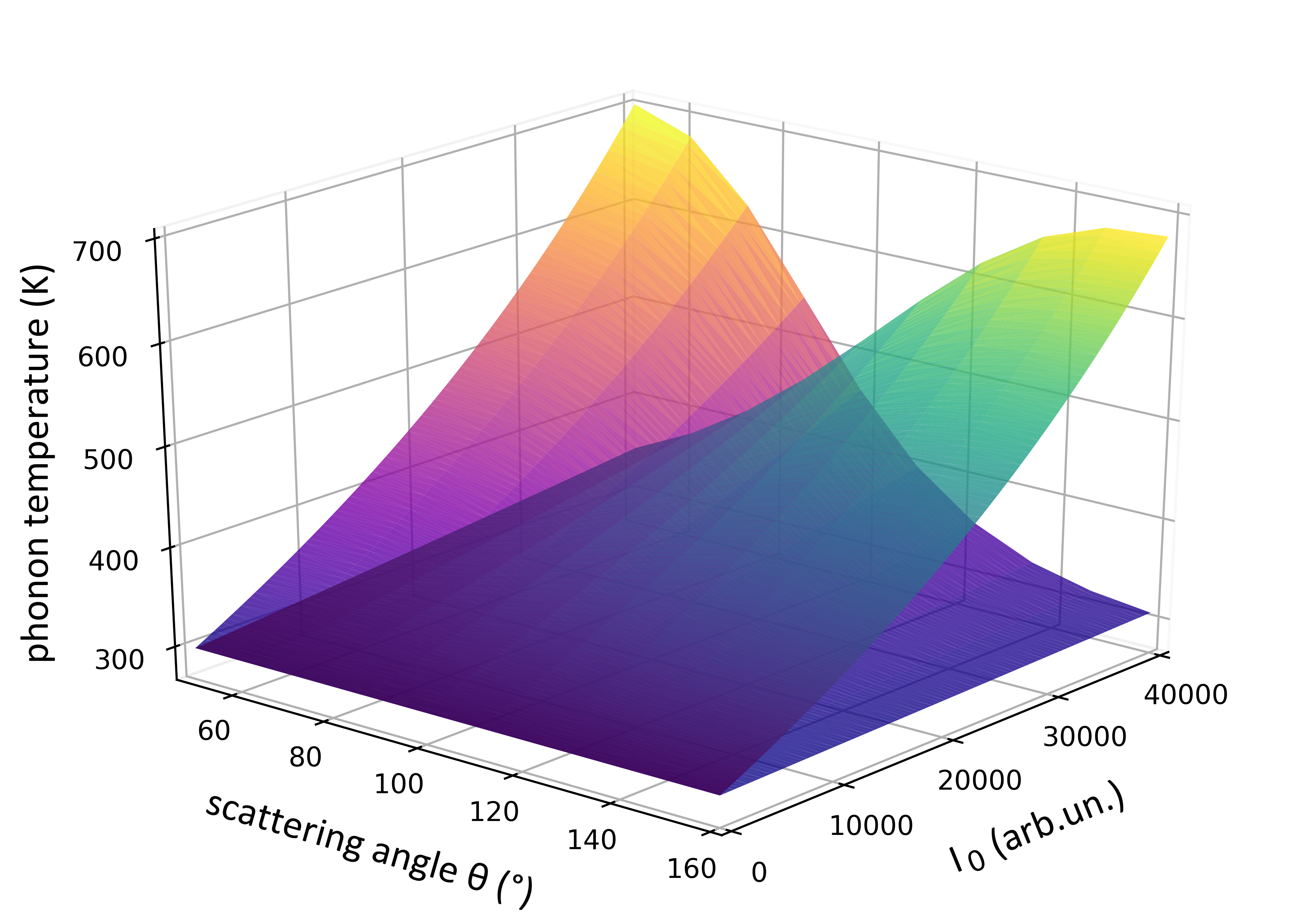}
    \caption{Phonon temperature versus $I_0$ and $\theta$ calculated from the scattering-modified phonon occupation number for the  LO (blue-to-yellow curve, increasing at growing $\theta$) and LA (purple-to-yellow curve) phonon modes. $\tilde{n}_{\bm{q}j} (t_f)$ is estimated by $R_{scat}$ data fitting for $\lambda_0^{(1)} = 4.08$ nm, $\epsilon_0 = $LV.}
    \label{phon-T}
\end{figure}

The last remark is on the final phonon state, having the characteristic of a coherent state with a specific change in the average occupation number. According to the model, quantitative information on the increase of $\tilde n_{\bm{q}j}(t)$ by scattering can be extracted from fitting as a byproduct. In Fig.\ref{phon-T}, we show the increase of the equivalent phonon temperature $T_{\bm{q}j}$, as estimated from $\tilde{n}_{\bm{q}j} (t_f)$. An increase of $T_{\bm{q}j}$ of the order of some hundreds K is estimated for both modes involved. The $T_{\bm qj}$ trends result almost independent of $\lambda_0$ , $\epsilon_0$.
It is important to observe that the equivalent temperature is just a way to quantify the growth of the phonon occupation number as it is not the thermodynamic temperature. Of course, at later times, once the anharmonic phonon-phonon interaction kicks in, the energy selectively deposited in such low-$q$ hot phonon populations is redistributed among all the vibrational (as well as electronic) degrees of freedom of the system, resulting in a much lower final equilibrium temperature. The different temperatures estimated for the LA and LO phonon populations reflect the selection rules for the modes at varying scattering geometry.

\section*{Conclusions}
In conclusion, we carried out a Thomson scattering experiment on HOPG exploiting the coherent, ultrashort EUV pulses of FERMI FEL. Here, we find out the evident, crucial role of quantum coherence, as the exponential growth of the scattered intensity upon increasing incident intensity is due to the non-zero commutator of the annihilation and creation operators of the radiation final coherent state. The nonlinearity here paralleling Dicke's theory of superradiance \cite{dicke, superradiant2} represents, to the best of our knowledge, the first direct laboratory observation of coherence of the scattered EUV radiation, also promoting low-energy coherent phonons in the material, as seen in the exponential trend of $d \sigma / d \Omega \vert_{ph}$. 
The QM description of the process provides a precise picture of the data, pointing to the way to obtain useful information on the specimen using a wide sampling of the available BZ volume. The present data already provide quantitative results regarding the determination of the absorption in bulk materials and the combined analysis of absorption and scattering factors in the energy region across the absorption edge, other than the study of the anisotropy of sound propagation in graphite.

Our results show that the described framework is the basis for conceiving novel EUV Thomson scattering experiments, opening promising routes for material investigations. These experiments could provide new information on the properties of condensed matter systems: firstly, they can give access to basic data, like absorption coefficient in the EUV, scattering factors in crossing the low energy absorption edges, and low-$q$ phonon dispersion curves in thin films. Secondly, they allow exploring exotic effects, like those produced by superradiant processes triggered by the bright and coherent EUV FEL pulses, combined with the reduced scattering volume in reciprocal space.

\bibliographystyle{unsrt}
%\bibliography{biblio}

\begin{thebibliography}{10}

\bibitem{heitler}
Walter Heitler.
\newblock {\em The quantum theory of radiation}.
\newblock Courier Corporation, 1984.

\bibitem{xray-theory}
Sunil~K Sinha.
\newblock Theory of inelastic x-ray scattering from condensed matter.
\newblock {\em Journal of Physics: Condensed Matter}, 13(34):7511, 2001.

\bibitem{astro1}
CF~Kennel and FV~Coroniti.
\newblock Confinement of the crab pulsar's wind by its supernova remnant.
\newblock {\em The Astrophysical Journal}, 283:694--709, 1984.

\bibitem{astro2}
DB~Wilson and MJ~Rees.
\newblock Induced compton scattering in pulsar winds.
\newblock {\em Monthly Notices of the Royal Astronomical Society},
  185(2):297--304, 1978.

\bibitem{coherence-X}
Vincent Moncrief.
\newblock Coherent states and quantum nonperturbing measurements.
\newblock {\em Annals of Physics}, 114(1-2):201--214, 1978.

\bibitem{SLAC1}
Paul Emma, R~Akre, J~Arthur, R~Bionta, C~Bostedt, J~Bozek, A~Brachmann,
  P~Bucksbaum, Ryan Coffee, F-J Decker, et~al.
\newblock First lasing and operation of an {\aa}ngstrom-wavelength
  free-electron laser.
\newblock {\em nature photonics}, 4(9):641--647, 2010.

\bibitem{FLASH}
J{\"o}rg Rossbach, Jochen~R Schneider, and Wilfried Wurth.
\newblock 10 years of pioneering x-ray science at the free-electron laser flash
  at desy.
\newblock {\em Physics reports}, 808:1--74, 2019.

\bibitem{SLAC-Xray}
LB~Fletcher, HJ~Lee, T~D{\"o}ppner, E~Galtier, B~Nagler, P~Heimann, C~Fortmann,
  S~LePape, T~Ma, M~Millot, et~al.
\newblock Ultrabright x-ray laser scattering for dynamic warm dense matter
  physics.
\newblock {\em Nature photonics}, 9(4):274--279, 2015.

\bibitem{PRL-Xray}
T~Ma, T~D{\"o}ppner, RW~Falcone, L~Fletcher, C~Fortmann, DO~Gericke, OL~Landen,
  HJ~Lee, A~Pak, J~Vorberger, et~al.
\newblock X-ray scattering measurements of strong ion-ion correlations in
  shock-compressed aluminum.
\newblock {\em Physical review letters}, 110(6):065001, 2013.

\bibitem{thom3}
HJ~Lee, P~Neumayer, J~Castor, T~D{\"o}ppner, RW~Falcone, C~Fortmann, BA~Hammel,
  AL~Kritcher, OL~Landen, RW~Lee, et~al.
\newblock X-ray thomson-scattering measurements of density and temperature in
  shock-compressed beryllium.
\newblock {\em Physical review letters}, 102(11):115001, 2009.

\bibitem{thom4}
C~Fortmann, HJ~Lee, T~D{\"o}ppner, RW~Falcone, AL~Kritcher, OL~Landen, and
  SH~Glenzer.
\newblock Measurement of the adiabatic index in be compressed by
  counterpropagating shocks.
\newblock {\em Physical Review Letters}, 108(17):175006, 2012.

\bibitem{thom5}
JC~Valenzuela, C~Krauland, D~Mariscal, I~Krasheninnikov, C~Niemann, T~Ma,
  P~Mabey, G~Gregori, P~Wiewior, AM~Covington, et~al.
\newblock Measurement of temperature and density using non-collective x-ray
  thomson scattering in pulsed power produced warm dense plasmas.
\newblock {\em Scientific reports}, 8(1):1--8, 2018.

\bibitem{quo-vadis}
Michael Bonitz, Zh~A Moldabekov, and TS~Ramazanov.
\newblock Quantum hydrodynamics for plasmas—quo vadis?
\newblock {\em Physics of Plasmas}, 26(9):090601, 2019.

\bibitem{thom-rev}
Siegfried~H Glenzer and Ronald Redmer.
\newblock X-ray thomson scattering in high energy density plasmas.
\newblock {\em Reviews of Modern Physics}, 81(4):1625, 2009.

\bibitem{ppscattering}
A~H{\"o}ll, Th~Bornath, L~Cao, T~D{\"o}ppner, S~D{\"u}sterer, E~F{\"o}rster,
  C~Fortmann, SH~Glenzer, G~Gregori, T~Laarmann, et~al.
\newblock Thomson scattering from near-solid density plasmas using soft x-ray
  free electron lasers.
\newblock {\em High Energy Density Physics}, 3(1-2):120--130, 2007.

\bibitem{eis}
Claudio Masciovecchio, Andrea Battistoni, Erika Giangrisostomi, Filippo
  Bencivenga, Emiliano Principi, Riccardo Mincigrucci, Riccardo Cucini,
  Alessandro Gessini, Francesco D'Amico, Roberto Borghes, et~al.
\newblock Eis: the scattering beamline at fermi.
\newblock {\em Journal of synchrotron radiation}, 22(3):553--564, 2015.

\bibitem{coherent-phon1}
Mariano Trigo, Matthias Fuchs, Jian Chen, MP~Jiang, Marco Cammarata, Stephen
  Fahy, David~M Fritz, Kelly Gaffney, Shambhu Ghimire, Andrew Higginbotham,
  et~al.
\newblock Fourier-transform inelastic x-ray scattering from time-and
  momentum-dependent phonon--phonon correlations.
\newblock {\em Nature Physics}, 9(12):790--794, 2013.

\bibitem{coherent-phon2}
Michael Kozina, M~Trigo, M~Chollet, JN~Clark, JM~Glownia, AC~Gossard,
  T~Henighan, MP~Jiang, H~Lu, A~Majumdar, et~al.
\newblock Heterodyne x-ray diffuse scattering from coherent phonons.
\newblock {\em Structural Dynamics}, 4(5):054305, 2017.

\bibitem{LiF}
C~Fasolato, F~Sacchetti, P~Postorino, P~Tozzi, E~Principi, A~Simoncig,
  L~Foglia, R~Mincigrucci, F~Bencivenga, C~Masciovecchio, et~al.
\newblock Ultrafast plasmon dynamics in crystalline lif triggered by intense
  extreme uv pulses.
\newblock {\em Physical Review Letters}, 124(18):184801, 2020.

\bibitem{FERMI}
Miltcho~B Danailov, Filippo Bencivenga, Flavio Capotondi, Francesco Casolari,
  Paolo Cinquegrana, Alexander Demidovich, Erika Giangrisostomi, Maya~P
  Kiskinova, Gabor Kurdi, Michele Manfredda, et~al.
\newblock Towards jitter-free pump-probe measurements at seeded free electron
  laser facilities.
\newblock {\em Optics express}, 22(11):12869--12879, 2014.

\bibitem{abs-sat1}
Linda Young, Elliot~P Kanter, Bertold Kraessig, Yongjin Li, AM~March, ST~Pratt,
  Robin Santra, SH~Southworth, Nina Rohringer, LF~DiMauro, et~al.
\newblock Femtosecond electronic response of atoms to ultra-intense x-rays.
\newblock {\em Nature}, 466(7302):56--61, 2010.

\bibitem{abs-sat2}
Hitoki Yoneda, Yuichi Inubushi, Makina Yabashi, Tetsuo Katayama, Tetsuya
  Ishikawa, Haruhiko Ohashi, Hirokatsu Yumoto, Kazuto Yamauchi, Hidekazu
  Mimura, and Hikaru Kitamura.
\newblock Saturable absorption of intense hard x-rays in iron.
\newblock {\em Nature communications}, 5(1):1--5, 2014.

\bibitem{abs-sat3}
Andrea Di~Cicco, Keisuke Hatada, Erika Giangrisostomi, Roberto Gunnella,
  Filippo Bencivenga, Emiliano Principi, Claudio Masciovecchio, and Adriano
  Filipponi.
\newblock Interplay of electron heating and saturable absorption in ultrafast
  extreme ultraviolet transmission of condensed matter.
\newblock {\em Physical Review B}, 90(22):220303, 2014.

\bibitem{dicke}
Robert~H Dicke.
\newblock Coherence in spontaneous radiation processes.
\newblock {\em Physical review}, 93(1):99, 1954.

\bibitem{superradiant2}
Michel Gross and Serge Haroche.
\newblock Superradiance: An essay on the theory of collective spontaneous
  emission.
\newblock {\em Physics reports}, 93(5):301--396, 1982.

\bibitem{phonon-lifetime}
Nicola Bonini, Michele Lazzeri, Nicola Marzari, and Francesco Mauri.
\newblock Phonon anharmonicities in graphite and graphene.
\newblock {\em Physical review letters}, 99(17):176802, 2007.

\bibitem{glauber}
Roy~J Glauber.
\newblock Coherent and incoherent states of the radiation field.
\newblock {\em Physical Review}, 131(6):2766, 1963.

\bibitem{Torres2017}
Theo Torres, Sam Patrick, Antonin Coutant, Maur{\'{\i}}cio Richartz, Edmund~W.
  Tedford, and Silke Weinfurtner.
\newblock Rotational superradiant scattering in a vortex flow.
\newblock {\em Nature Physics}, 13(9):833--836, June 2017.

\bibitem{Inouye1999}
S.~Inouye, A.~P. Chikkatur, D.~M. Stamper-Kurn, J.~Stenger, D.~E. Pritchard,
  and W.~Ketterle.
\newblock Superradiant rayleigh scattering from a bose-einstein condensate.
\newblock {\em Science}, 285(5427):571--574, July 1999.

\bibitem{Cong2016}
Kankan Cong, Qi~Zhang, Yongrui Wang, G.~Timothy Noe, Alexey Belyanin, and
  Junichiro Kono.
\newblock Dicke superradiance in solids [invited].
\newblock {\em Journal of the Optical Society of America B}, 33(7):C80, May
  2016.

\bibitem{Masson2022}
Stuart~J. Masson and Ana Asenjo-Garcia.
\newblock Universality of dicke superradiance in arrays of quantum emitters.
\newblock {\em Nature Communications}, 13(1), April 2022.

\bibitem{TMD}
Sajedeh Manzeli, Dmitry Ovchinnikov, Diego Pasquier, Oleg~V Yazyev, and Andras
  Kis.
\newblock 2d transition metal dichalcogenides.
\newblock {\em Nature Reviews Materials}, 2(8):1--15, 2017.

\bibitem{FERMI2}
E~Allaria, Roberto Appio, L~Badano, WA~Barletta, S~Bassanese, SG~Biedron,
  A~Borga, E~Busetto, D~Castronovo, P~Cinquegrana, et~al.
\newblock Highly coherent and stable pulses from the fermi seeded free-electron
  laser in the extreme ultraviolet.
\newblock {\em Nature Photonics}, 6(10):699--704, 2012.

\bibitem{ref353}
Jay~A Brandes, George~D Cody, Douglas Rumble, Paul Haberstroh, Sue Wirick, and
  Yves Gelinas.
\newblock Carbon k-edge xanes spectromicroscopy of natural graphite.
\newblock {\em Carbon}, 46(11):1424--1434, 2008.

\bibitem{graphite-phon-n}
R~Nicklow, N~Wakabayashi, and HG~Smith.
\newblock Lattice dynamics of pyrolytic graphite.
\newblock {\em Physical Review B}, 5(12):4951, 1972.

\bibitem{graphite-phon-X}
J~Maultzsch, S~Reich, C~Thomsen, H~Requardt, and P~Ordej{\'o}n.
\newblock Phonon dispersion in graphite.
\newblock {\em Physical review letters}, 92(7):075501, 2004.

\bibitem{graphite-phon-X2}
M~Mohr, J~Maultzsch, E~Dobard{\v{z}}i{\'c}, S~Reich, I~Milo{\v{s}}evi{\'c},
  M~Damnjanovi{\'c}, A~Bosak, M~Krisch, and C~Thomsen.
\newblock Phonon dispersion of graphite by inelastic x-ray scattering.
\newblock {\em Physical Review B}, 76(3):035439, 2007.

\bibitem{elastic-const-graphite}
G~Savini, YJ~Dappe, Sven {\"O}berg, J-C Charlier, MI~Katsnelson, and
  A~Fasolino.
\newblock Bending modes, elastic constants and mechanical stability of
  graphitic systems.
\newblock {\em Carbon}, 49(1):62--69, 2011.

\bibitem{NIST}
Christopher~T Chantler.
\newblock Theoretical form factor, attenuation, and scattering tabulation for
  z= 1--92 from e= 1--10 ev to e= 0.4--1.0 mev.
\newblock {\em Journal of Physical and Chemical Reference Data}, 24(1):71--643,
  1995.

\bibitem{29a}
John~C Parker and RH~Pratt.
\newblock Validity of common assumptions for anomalous scattering.
\newblock {\em Physical Review A}, 29(1):152, 1984.

\bibitem{pines}
D~Pines and P~Nozieres.
\newblock The theory of quantum liquids. wa benjamin, inc.
\newblock {\em New York}, 1966.

\bibitem{TDS}
BTM Willis.
\newblock Thermal diffuse scattering of x-rays and neutrons.
\newblock {\em International Tables for Crystallography}, B; 4.1(1):484--491,
  2010.

\bibitem{graphite-theory}
John~P Perdew, Adrienn Ruzsinszky, G{\'a}bor~I Csonka, Oleg~A Vydrov, Gustavo~E
  Scuseria, Lucian~A Constantin, Xiaolan Zhou, and Kieron Burke.
\newblock Restoring the density-gradient expansion for exchange in solids and
  surfaces.
\newblock {\em Physical Review Letters}, 100(13):136406, 2008.

\bibitem{elk-code}
Elk code.
\newblock {http://elk.sourceforge.net/}, 2021.
\newblock The exchange and correlation potential from ref.
  \cite{graphite-theory} was used. The intrinsic disorder of the HOPG planes
  about the c-axis was accounted for by calculating the dispersion relations
  along the $\Gamma_M$ and $\Gamma_K$ directions. Both data are reported in
  Figure \ref{fig4_vs-q}a.

\end{thebibliography}

\end{document}